\documentclass[10pt,conference]{IEEEtran}

\usepackage{xr}

\usepackage{todonotes}

\usepackage{subcaption}
\usepackage{acronym}

\usepackage{graphicx}
\usepackage{microtype}
\usepackage{url}

\usepackage{mathtools}
\usepackage{amsmath}
\usepackage{cleveref}
\usepackage{amssymb}
\usepackage[altpo]{backnaur}
\usepackage{listings}

\usepackage{amsthm}

\newtheoremstyle{bolditalicdef} 
  {0.7\topsep}   
  {0.7\topsep}   
  {\normalfont} 
  {}          
  {\bfseries\itshape} 
  {.}         
  { }         
  {\thmname{#1}\thmnumber{ #2}\thmnote{ (#3)}} 

\theoremstyle{bolditalicdef}
\newtheorem{definition}{Definition}

\usepackage{multirow}
\usepackage{tabularx,adjustbox,booktabs}
\usepackage{siunitx}
\usepackage{dcolumn}

\usepackage{array}

\usepackage[english]{babel}

\usepackage[]{xcolor}
\usepackage{colortbl}

\usepackage{csquotes}
\usepackage{csvsimple}
\usepackage[para]{footmisc}
\usepackage{xcolor}

\usepackage{caption}

\usepackage[breakable, skins,xparse]{tcolorbox}

\usepackage{tikz}
\usetikzlibrary{shapes.geometric, shapes,arrows, graphs, matrix}

\usepackage{pifont}

\newcolumntype{L}{>{\phantom{$-$}}l}       

\newcommand\restr[2]{{
  \left.\kern-\nulldelimiterspace 
  #1 
  \vphantom{\big|} 
  \right|_{#2} 
  }}
 
\usepackage[numbers]{natbib}

\usepackage{listings}

\usepackage[ruled,vlined,linesnumbered,inoutnumbered,noend]{algorithm2e}

\crefname{chapter}{Chapter}{Chapters}
\crefname{section}{Section}{Sections}
\crefname{subsection}{Subsection}{Subsections}
\crefname{figure}{Figure}{Figures}
\crefname{table}{Table}{Tables}
\crefname{listing}{Listing}{Listings}

\crefname{definition}{Definition}{Definitions}

\newcommand{\xaxis}{$x$-axis\xspace}
\newcommand{\yaxis}{$y$-axis\xspace}

\newenvironment{rqquestion}[1]{
    \begin{tcolorbox}
    [arc=0pt,outer arc=0pt, boxrule=0pt, top=2pt, bottom=2pt, left=3pt, right=3pt, breakable,  sharpish corners,enhanced, drop lifted shadow] 
    \textbf{{#1}:}
}{
    \end{tcolorbox}
}

\tcbset{
    mydefault/.style={
        colback=white, 
        colframe=black, 
        boxrule=0.5pt, 
        sharp corners, 
        enhanced,  
        width=\linewidth
    }
}

\newcommand{\welchwop}[2]{$t(#2)=#1$}

\DeclareRobustCommand{\rchi}{{\mathpalette\irchi\relax}}
\newcommand{\irchi}[2]{\raisebox{\depth}{$#1\chi$}}

\newcommand{\suparagraph}[1]{\vskip -2ex \leavevmode\unskip\\\noindent\textbf{#1.}}

\newcommand{\constant}[1]{${#1}$\xspace}

\newcommand{\numParticipantsTotal}{\constant{19}}
\newcommand{\numSessionsTotal}{\constant{13}}
\newcommand{\numSessionsCopilot}{\constant{7}}
\newcommand{\numSessionsPP}{\constant{6}} 

\newcommand{\meanExpAbs}{\constant{6.21}}

\newcommand{\meanExpCopilotCopilot}{\constant{4.43}}

\newcommand{\meanExpCopilotPP}{\constant{1.33}}

\newcommand{\corrAbsRelExp}{\constant{0.62}}

\newcommand{\numEpTotalCopilot}{\constant{126}} 
\newcommand{\numEpTotalPP}{\constant{210}} 

\newcommand{\numEpMeanCopilot}{\constant{18.0}} 
\newcommand{\numEpMeanPP}{\constant{35.0}}

\newcommand{\numEpFstQuCopilot}{\constant{15}}
\newcommand{\numEpMedianCopilot}{\constant{17}}
\newcommand{\numEpThdQuCopilot}{\constant{22}}
\newcommand{\numEpFstQuPP}{\constant{28}}
\newcommand{\numEpMedianPP}{\constant{33.5}}
\newcommand{\numEpThdQuPP}{\constant{35}}

\newcommand{\wtNumEpT}{-3.38}

\newcommand{\wtNumEpDF}{8.15}

\newcommand{\lenEpMeanCopilot}{\constant{9.4}}
\newcommand{\lenEpMeanPP}{\constant{14.4}} 
\newcommand{\lenEpMedianCopilot}{\constant{6.5}}
\newcommand{\lenEpMedianPP}{\constant{8}}

\usepackage{ifthen}

\newboolean{changed}
\setboolean{changed}{false}
\newcommand{\changed}[1]{\ifthenelse{\boolean{changed}}{\color{red}#1\color{black}}{\color{black}#1\color{black}}}

\newboolean{cut}
\setboolean{cut}{true}
\newcommand{\cut}[2]{\ifthenelse{\boolean{cut}}{\color{yellow}#1\color{black}}{\color{black}#2\color{black}}}

\usepackage{svg}
\usepackage{float}
\definecolor{darkblue}{rgb}{0,0,.75}
\definecolor{eminence}{RGB}{108,48,130}
\definecolor{lapis}{HTML}{53ABFF}

\newcommand{\fname}[1]{{\scshape{{#1}}}}

\newcommand{\fterm}[1]{\emph{{#1}}}

\usepackage{etoolbox}
\robustify\acl
\DeclareRobustCommand{\a}[1]{\acl*{#1}\xspace}
\newcommand{\A}[1]{\Acl*{#1}\xspace}

\acrodef{IDE}{Integrated Development Environment}
\acrodefplural{IDE}[IDEs]{Integrated Development Environments}
\acrodef{UML}{Unified Modeling Language}
\acrodef{GPT}{generative pre-trained transformer}
\acrodef{LLM}{large language model}

\acrodef{github}{\fname{GitHub}}
\acrodef{copilot}{\acl*{github} \fname{Copilot}}
\acrodef{todoscode}{\textsf{$\#$TODO}}
\acrodef{openai}{\fname{OpenAI}}
\acrodef{chatgpt}{\fname{ChatGPT}}
\acrodef{whisper}{\fname{Whisper}}
\acrodef{codex}{\fname{Codex}}
\acrodef{microsoft}{\fname{Microsoft}}
\acrodef{office}{\acl*{microsoft} \fname{Office}}
\acrodef{java}{\fname{Java}}
\acrodef{python}{\fname{Python}}
\acrodef{c}{\fname{C}}
\acrodef{js}{\fname{JavaScript}}
\acrodef{leetcode}{\fname{LeetCode}}
\acrodef{vs}{\fname{Visual Studio}}
\acrodef{vscode}{\fname{Visual Studio Code}}
\acrodef{jetbrains}{\fname{JetBrains}}
\acrodef{pycharm}{\fname{PyCharm}}
\acrodef{true}{\fname{true}}
\acrodef{false}{\fname{false}}

\acrodef{kt}{knowledge transfer}
\acrodef{pp}{pair programming}
\acrodef{pps}{Pair Programming Setting}
\acrodef{humancopilot}{human--AI pair programming}
\acrodef{humanhuman}{human--human pair programming}

\begin{document}
\title{From Developer Pairs to AI Copilots:\\ A Comparative Study on Knowledge Transfer}

\author{
    Alisa Welter\textsuperscript{1},
    Niklas Schneider\textsuperscript{1},
    Tobias Dick\textsuperscript{1},
    Kallistos Weis\textsuperscript{1},
    Christof Tinnes\textsuperscript{1,2},
    Marvin Wyrich\textsuperscript{1},
    Sven Apel\textsuperscript{1}
    \\
    \textsuperscript{1}Saarland University, Saarbrücken, Germany \\
    \textsuperscript{2}Siemens AG, München, Germany \\
    \{welter, nschneider, tdick, kallistos, wyrich, apel\}@cs.uni-saarland.de, christof.tinnes@siemens.com
}










\maketitle


\begin{abstract}
\A{kt} is fundamental to human collaboration and is therefore common in software engineering. Pair programming is a prominent instance.
With the rise of AI coding assistants, developers now not only work with human partners but also, as some claim, with \emph{AI pair programmers}. 
Although studies confirm knowledge transfer during human pair programming, its effectiveness with AI coding assistants remains uncertain.

To analyze \a{kt} in both human–human and human–AI settings, we conducted an empirical study where developer pairs solved a programming task without AI support, while a separate group of individual developers completed the same task using the AI coding assistant \a{copilot}.
We extended an existing \a{kt} framework and employed a semi-automated evaluation pipeline to assess differences in \a{kt} episodes across both settings. We found a similar frequency of successful knowledge transfer episodes and overlapping topical categories across both settings.
Two of our key findings are that developers tend to accept \a{copilot}'s suggestions with less scrutiny than those from human pair programming partners, but also that \a{copilot} can subtly remind developers of important code details they might otherwise overlook.

\end{abstract}


\IEEEdisplaynontitleabstractindextext
\section{Introduction}\label{sec:introduction}

\A{pp} has been shown to be an efficient technique, yielding higher quality and more readable code compared to individual efforts~\cite{dybaa2007two, jones2013backseatdriver}. 
While pair programming can reduce errors, its lower time efficiency~\cite{dybaa2007two} often leads to it being underused in practice.
Meanwhile, advances in AI have begun to reshape coding practices.
The advent of \a{copilot}~\cite{chen2021copilot} in late 2021, followed by other coding assistants, has noticeably changed how code is written~\cite{vaithilingam2022expectation, ziegler2024measuring}.
\acl*{github}, for example, reports a growing number of companies adopting \a{copilot}\footnote{\url{https://github.com/features/copilot/}} to increase productivity, claiming that one in three Fortune 500 companies now employs it.\footnote{\url{https://www.microsoft.com/investor/reports/ar24/index.html}} Similarly, Google's CEO Sundar Pichai notes that over a quarter of new code at Google is now AI-generated.\footnote{\url{https://blog.google/inside-google/message-ceo/alphabet-earnings-q3-2024/}}

This emerging technical landscape has sparked the idea of \enquote{AI pair programmers}, that is, replacing one of the two human developers in a pair programming setting with an AI coding assistant. A key question, however, is whether an AI coding assistant can effectively replace the second human while preserving the core benefits of traditional pair programming.
In particular, one significant long-term benefit of pair programming is knowledge transfer~\cite{zieris2014knowledge}, which is the focus of our research.

\A{kt} frequently happens during \a{pp}, as developers continuously discuss issues and their solutions as they progress~\cite{jones2013backseatdriver, zieris2014knowledge, kuttal2021tradeoffs}.
This creates learning opportunities through knowledge exchange and valuable discussions that contribute to solving tasks or resolving challenges. For instance, a developer might explain some domain logic, guide their peer through a complex algorithm, or simply demonstrate a new keyboard shortcut in their coding environment.
It is important to note that knowledge transfer can occur subtly and does not always result from direct questions and answers. It can emerge naturally during collaboration, such as when developers proactively share insights or demonstrate best practices.

In the literature, different views on \a{kt} in pair programming have been explored~\cite{zieris2014knowledge, plonka2015knowledge}. 
In general, collaborative techniques such as discussions and peer teaching have been shown to deepen learners' understanding of concepts and problems compared to working alone~\cite{stigmar2016peer, hattie2009visible}. 
With the idea of AI pair programming, a second human is not always involved in such pair activities.
\mbox{\citet{kuttal2021tradeoffs}} studied a digital agent with a simulated face that engaged programmers through questions and suggestions, but its limited reasoning hindered effective knowledge exchange.
\mbox{\citet{imai2022github}} investigated changes in productivity and code quality in pair programming with \a{copilot} compared to a human partner.
In their study, \a{copilot} enhanced productivity but reduced code quality, as indicated by an increase in deleted lines.

The broader picture is that, while existing research has either focused on understanding and improving \a{humanhuman} or on using agents to enhance efficiency, only little attention has been paid to directly comparing knowledge transfer between \a{humanhuman} and pair programming with AI coding assistent.
We therefore set out to investigate \a{kt} in programming with AI coding assistants, particularly \a{copilot}, and compare it to traditional \a{pp}.
To measure \a{kt}, we devise a \a{kt} framework based on the works of~\mbox{\citet{zieris2014knowledge}} and~\mbox{\citet{kuttal2021tradeoffs}}.
Based on this framework, we establish a semi-automated pipeline to evaluate the data collected from our empirical study, in which 
a total of \numParticipantsTotal developers either code as pairs or use \a{copilot} on a programming problem that required participants to develop algorithms and integrate them into a cohesive project framework.

Our results show that knowledge transfer occurs in both \a{humanhuman} and \a{humancopilot}.
\a{copilot} can even subtly provide helpful reminders, such as suggesting missing database commits.
While human pairs engage in more frequent but sometimes distracting exchanges, \a{copilot} promotes more focused interactions---though users tend to trust its suggestions more readily, raising questions about critical engagement.

In summary, our contributions are as follows:  

\begin{itemize}

\item We unify and extend the perspectives of~\citet{zieris2014knowledge} and~\citet{kuttal2021tradeoffs} into a single framework for \a{kt} in the context of AI coding assistants.
\item We conduct an empirical study comparing \a{kt} in AI-assisted and traditional pair programming, revealing key differences between the two.
\item Drawing on our findings, we discuss how combining both approaches can enhance development efficiency and focus while preserving the benefits of human collaboration.
\end{itemize}

\section{Background and Related Work}\label{sec:relatedworkNEW}

This section reviews prior work on knowledge transfer, pair programming, and AI assistants, forming the conceptual basis of our study alongside the framework in Section~\ref{sec:knowledgetransfer}.

\subsection{Knowledge Transfer}
\A{kt} plays a crucial role, for example, in education \cite{bada2015constructivism, nokes2015better, johnson2007state, stigmar2016peer, goldschmid1976peer}. 
Numerous studies indicate that peers working together often outperform individuals on certain tasks through discussions, explanations, and the sharing of ideas \cite{nokes2015better, johnson2007state}. 
However, understanding the nature of these processes requires a profound analysis. 
According to constructivist learning theory, knowledge cannot be transmitted directly as it is not a tangible entity \cite{liu2005vygotsky, bada2015constructivism}. 
Instead, individuals must explore and construct knowledge through experiences and by resolving conflicts with their existing knowledge \cite{piaget1980psychogenesis}. 
In general, \a{kt} involves actions and mechanisms that facilitate this construction process, such as explanations and discussions that can take place in pair programming.

\subsection{Pair Programming}
\emph{Pair programming} is a technique in which two developers work collaboratively at a single computer~\cite{beck1999embracing}.
We use this technique as a baseline for our study, as it has been researched extensively~\cite{arisholm2007evaluating, begel2008pair, dybaa2007two}. 
\mbox{\citet{zieris2014knowledge}} explore \a{kt} in pair programming and propose a framework detailing how \a{kt} processes are structured. They identify factors that facilitate effective \a{kt} \cite{zieris2014knowledge, zieris2016observations, zieris2020qualitative} with the overarching goal of providing practical advice for programmers.
Dybå et al.'s meta-analysis \cite{dybaa2007two} examines the effectiveness of pair programming by reviewing the results of $15$ studies focusing on various aspects such as work speed, effort, and quality: 
With pair programming, tasks are completed faster and software quality improves.
However, pair programming is more expensive than individual programming due to the increased number of person-hours required.
Additionally, the benefit of pair programming declines with programmer expertise and task complexity~\cite{dybaa2007two}.

\subsection{AI Coding Assistants}

While pair programming is traditionally between two humans, an AI coding assistant can take on tasks typically handled by a human partner. \a{copilot} is a popular example, powered by the \ac{LLM} \a{codex}\cite{chen2021copilot}. 
Released by \a{microsoft} in 2021, it was one of the first AI assistants to be integrated into \acp{IDE} via plugins. At the time of this study, only \a{codex} was available as an underlying model, although more recent versions of \a{copilot} now support various \acp{LLM}.
In this study, we differentiate between \emph{\a{humanhuman}}, where two developers work together on one computer, and \emph{\a{humancopilot}}, where a developer collaborates with an AI coding assistant, \a{copilot}.

Several studies have focused on the performance of \a{copilot} regarding different aspects, mostly correctness.
\mbox{\citet{nguyen2022empirical}} use \a{leetcode}\footnote{\url{https://leetcode.com/}} questions as prompts for \a{copilot} to solve programming tasks in four different languages. 
Using \a{leetcode} tests and code complexity metrics, they find that correctness depends strongly on the programming language, with \a{java} performing best, and \a{copilot}'s code being highly comprehensible as indicated by low complexity values. Peng et al.~\cite{peng2023developerproductivity} observe an improvement in the time needed to solve programming tasks when using AI assistants in a controlled experiment.

Concerning the analysis of pair programming in a setting where one of the partners is replaced with a digital agent,
Kuttal et al.~\cite{kuttal2021tradeoffs} explore an approach where the agent, represented by a simulated face on the screen, interacts with the human programmer by prompting questions and offering ideas. 
They find no significant difference in the typical advantages of pair programming, such as productivity, quality, and self-efficacy, maintaining the benefits of the traditional approach.
Participants trust and accept the simulated face on the screen as a partner, yet the agent struggles to explain logic or contribute original ideas, impairing effective knowledge exchange. The study focuses on replicating human-like features, such as conversation and a simulated avatar, whereas our approach emphasizes code generation while offering the human programmer an experience similar to working with a human partner.
In a study conducted by Imai~\cite{imai2022github}, participants are asked to implement a Minesweeper game in Python through pair programming, with \a{copilot} partially serving as a programming partner. The objective of the study is to measure productivity and code quality. 
The study measures productivity by lines added and quality by deletions. \a{copilot} users are most productive but produce lower-quality code.

Ma et al.~\cite{ma2023pAIrprogramming} compare \a{humanhuman} and \a{humancopilot} across effectiveness metrics (learning, cost, collaboration, communication) and call for more \a{humancopilot} research to fully leverage AI in pair programming. Barke et al.~\cite{barke2023grounded} perform a study to identify two key interaction modes with \a{copilot}: In the acceleration mode, programmers have a clear plan and use \a{copilot} as an autocomplete tool, quickly accepting suggestions without disrupting their workflow. In exploration mode, programmers are unsure how to solve a problem and use \a{copilot} to explore various approaches~\cite{barke2023grounded}.

\changed{\citet{stray2025generative} interview developers and report that AI assistants were said to improve efficiency and reduce stress during individual programming, giving developers more time to focus on solving complex tasks. However, the also noted that AI assistants were rarely used in \a{humanhuman}.

\citet{fan2025impact} conduct a quasi-experiment with students and find that \a{humancopilot} improves motivation, reduced anxiety, and outperforms individual work, while \a{humanhuman} leads to the highest sense of collaboration and social presence~\cite{fan2025impact}.}

To summarize, previous research focused on understanding and improving \a{humanhuman} or using agents to boost efficiency.
Only little attention has been given to directly comparing knowledge transfer between \a{humanhuman} and \a{humancopilot}, which is however imperative to understand the impact of the increasing use of AI as a programmer.
\changed{\section{A Knowledge Transfer Framework for Human--AI Pair Programming}}
\label{sec:knowledgetransfer}

To compare \a{kt} in \a{humanhuman} and \a{humancopilot}, we devise a \a{kt} framework based on~\mbox{\citet{zieris2014knowledge}} and~\mbox{\citet{kuttal2021tradeoffs}}, ensuring comparability between the two settings. We analyze several metrics, as well as the content, and success of \a{kt} episodes, as described below. To define \a{kt}, we first define a knowledge gap.

\begin{definition} [Knowledge gap] A \emph{Knowledge gap} is a disparity between the information or understanding a person currently possesses and the knowledge a pair member considers relevant for their software development context including their current task and the software project it is embedded in~\cite{zieris2020qualitative}.
\end{definition}

Intuitively, a knowledge gap can only be bridged through the assimilation of new information, thereby leading to our definition of knowledge transfer.

\begin{definition}[Knowledge transfer]
    Any attempt to close a perceived knowledge gap by exchanging existing knowledge or building new knowledge is considered \emph{\a{kt}}~\cite{zieris2020qualitative}.
\end{definition}

The key aspects of \a{kt} according to \mbox{\citet{zieris2014knowledge}} in the context of \a{pp} are also applicable for \a{humancopilot}.~\mbox{\citet{zieris2014knowledge}} focus on the structure of \a{kt}, with which we will start.

In our framework, conversations are made of utterances. 
Since measuring knowledge transfer in individual utterances is impractical, we group related utterances into episodes.

\begin{definition}[Episode]
    An \fterm{episode} is defined as a sequence of related utterances focused on a single topic,  initiated either by a person explicitly expressing a need for knowledge or a person sharing knowledge they possess with others.
    \label{episodedef}
\end{definition}

Since an episode is initiated based on a perceived knowledge gap rather than an objectively existing one, a new episode may occur even when no actual knowledge gap is present. 

To allow statistical characterization of episodes, we define the \fterm{length} and \fterm{depth} of an episode. Intuitively, the length of an episode can be measured by counting the utterances it contains.

\begin{definition}[Length of an episode]
    The \fterm{length of an episode} is the number of utterances that belong to it, including the utterances belonging to episodes layered within it.
    \label{episodelengthdef}
\end{definition}
Episodes can be layered if a new topic is introduced before the current episode is complete. 
This layering requires a return to the topic of the initial episode. To formalize this structure, we define the concept of depth within episodes.

\begin{definition}[Depth of an episode]
    The \fterm{depth of an episode} refers to the total number of episodes that contain the given episode, with the top-level episode having a depth of one.
    \label{episodedepthdef}
\end{definition}

In other words, the depth of an episode refers to its nesting level within other episodes. A top-level episode, not nested within any other episode, is assigned a depth of one. Each time an episode is nested within another, depth increases by one.

Consider the following example:
A developer is using a new API and asks their partner for help, initiating an episode focused on API usage. While following their partner's explanation, a missing dependency causes an error, prompting a nested episode on installing it. Once the dependency issue is resolved, the discussion returns to the initial topic of using the API.
In this example, the length of the episode on API usage includes all utterances, including those from the inner dependency episode, while the inner episode only contains its own utterances. The depth of the inner episode is $2$, as it is enclosed within the episode on API usage, which has a depth of $1$. 


\suparagraph{Finish Types}
\mbox{\citet{zieris2014knowledge}} define attributes to characterize \a{kt} episodes. 
We focus on how an episode ends—its \emph{finish type} (\cref{tab:finish-types}).

\begin{table*}
    \centering
    \caption{Definition of knowledge transfer finish types, based on \mbox{\citet{zieris2014knowledge}}
    } 
    \begin{tabularx}{0.95\textwidth}{lp{8cm}X}
        \toprule
        \textbf{Finish Type} & \textbf{Definition} & \textbf{Example} \\
        \midrule
        \textsc{Assimilation} & The customer fully understands and internalizes the knowledge. & \emph{``I get it''} or rephrasing the explanation in their own words. \\
        \textsc{Trust} & The customer accepts the contribution without full understanding, relying on the partner’s expertise. & \emph{``Fine with me''} or \emph{``If Copilot says so, it’ll be right.''} \\
        \textsc{Gave up} & The attempt to close the knowledge gap is abandoned due to frustration or complexity. & \emph{``This is too complicated, never mind.''} \\
        \textsc{Lost sight} & The episode ends because the pair gets sidetracked, and the original goal is no longer pursued. & \emph{[Starts something unrelated while leaving the question unanswered]} \\
        \textsc{Unnecessary} & The knowledge is deemed irrelevant or redundant. & \emph{``I don’t think this is important right now.''} \\
        \bottomrule
    \end{tabularx}
    \label{tab:finish-types}
\end{table*}

Based on the four concrete finish types that Zieris and Prechelt introduce (i.e., \textsc{Transferred}, \textsc{Gave up}, \textsc{Lost sight}, and \textsc{Unnecessary}), we define five finish types of episodes in our framework. In particular, we split the finish type \emph{\textsc{Transferred}} into two finer-grained finish types: \emph{\textsc{Assimilation}} and \emph{\textsc{Trust}}.
To define these finish types, we first define the concept of a \fterm{customer}.

\begin{definition}[Customer]
     The person who is in need of knowledge is called \fterm{customer}~\cite{zieris2014knowledge}.
     \label{customerdef}
\end{definition}
Using \Cref{customerdef}, we define the finish types in our framework.
We start with the two finish types corresponding to \textsc{Transferred}, both of which describe situations in which the customer receives knowledge.

\begin{definition}[\textsc{Assimilation}]
    The episode concludes with the customer fully understanding and internalizing the knowledge shared by the partner.
    Knowledge transfer is complete and effective.
\end{definition}

\textsc{Assimilation} would include saying something like \enquote{I get it} because the customer understands the reasoning and agrees based on their own context. Instead of resulting in a short affirmative answer, transferred knowledge can also be rephrased by the customer.

\begin{definition}[\textsc{Trust}]
    The episode ends with the customer accepting the partner’s contribution without fully understanding it, relying on the partner’s perceived expertise or correctness.
    Knowledge is transferred but not deeply comprehended.
\end{definition}

An example of the \textsc{Trust} finish type would be
the statement \enquote{Fine with me}, which reflects acceptance of the information based on confidence in the source, rather than through personal understanding or critical engagement. 
Since genuine understanding is essential for applying knowledge effectively, we distinguish \textsc{Trust} from \textsc{Assimilation} even though such episodes can also be considered to represent a successful outcome.

However, knowledge has still been received even if not processed.
The next three finish types are concerned with scenarios where \a{kt} is not completed successfully.

\begin{definition}[\textsc{Gave up}]
    The episode ends due to the customer or pair abandoning the effort to bridge the knowledge gap, often from frustration, complexity, or lack of progress.
    No knowledge transfer occurs.
\end{definition}

In this case, a customer might express frustration with phrases like \enquote{This is too complicated, never mind}, signaling disengagement from the knowledge exchange. This outcome is undesirable, as it results in no meaningful transfer, leaving the knowledge gap unresolved.

\begin{definition}[\textsc{Lost sight}]
    The episode terminates because the pair diverges from the original knowledge transfer goal, becoming sidetracked or overwhelmed, leaving the knowledge gap unresolved.
\end{definition}

\textsc{Lost sight} reflects a situation in which either the customer or their partner gets sidetracked. This can result in the start of a new episode, leaving the current knowledge transfer unfinished and the knowledge gap unresolved.

\begin{definition}[\textsc{Unnecessary}]
    The episode ends when the pair recognizes the knowledge in question is irrelevant or redundant to the task, terminating \a{kt} as it is deemed non-essential.
\end{definition}

For instance, a customer might say \enquote{I don't think this is important right now,} reflecting a situation in which the current knowledge transfer is no longer the priority. In such cases, the knowledge gap persists because the focus has shifted away from the original objective.


\suparagraph{Topic Types} \mbox{\citet{zieris2014knowledge}} propose that every utterance should be labeled with an information type.
The overall information type of an episode is then derived from the most frequently occurring type among its individual utterances.
Regarding the information type, \mbox{\citet{zieris2014knowledge}} do not suggest a concrete classification scheme. 
In contrast, \mbox{\citet{kuttal2021tradeoffs}} explicitly address the content aspect of \a{kt}.
During their studies, they identified six distinct topics frequently discussed by developers (see \cref{tab:topics}).

\begin{table}[H]
    \centering
    \caption{Definition of knowledge transfer topic types, based on \mbox{\citet{kuttal2021tradeoffs}}}
    
    \begin{tabularx}{\linewidth}{lp{3.5cm}X}
        \toprule
        \textbf{Topic type} & \textbf{Definition} & \textbf{Example} \\
        \midrule
        \textsc{Tool} & Knowledge about the IDE or how to use the tool. & \emph{``There's a key bind for going back a tab''} \\
        \protect\textsc{Program} & Knowledge about the programming language itself or it's syntax. &  \emph{``I could also just aggregate the chars. I think they work like strings in that way.''} \\
        \protect\textsc{Bug} & An error in the code. &  \emph{``I think we are missing code in that function.''} \\
        \protect\textsc{Code} & The code itself; i.e., what they are programming &  \emph{``So I'm gonna need to use a for loop as well.''} \\
        \protect\textsc{Domain} & The task (in case of \cite{kuttal2021tradeoffs}: implement tic-tac-toe game) &  \emph{``The goal is to get three of ... marker ... in a row.''} \\
        \protect\textsc{Technique} & About the techniques being used (i.e., test driven development, pair programming) &  \emph{``So if it's test driven development, I want to start by writing a test, I think.''} \\
        \bottomrule
    \end{tabularx}
    \label{tab:topics}
\end{table}
\changed{
Instead of using the information type attribute defined by \mbox{\citet{zieris2014knowledge}}, we thus classify utterances using topic types based on the classes established by \mbox{\citet{kuttal2021tradeoffs}}.
Although this list of classes cannot be exhaustive, it provides a structured overview of common topics discussed in \a{kt} episodes.
Specifically, the episode's topic type  is determined by the most frequently occurring topic type among its constituent utterances.}\@
This approach aligns with the models proposed by both \mbox{\citet{zieris2014knowledge}} and \mbox{\citet{kuttal2021tradeoffs}}, as both sets of categories can apply consistently at the utterances level, ensuring compatibility between the two models.

In summary, our knowledge transfer framework facilitates a comprehensive analysis of knowledge transfer for both our \a{humanhuman} and \a{humancopilot} settings.

\section{Methodology}\label{sec:methodology}

\begin{figure*}[ht]
    \centering
    \includegraphics[width=0.8\textwidth]{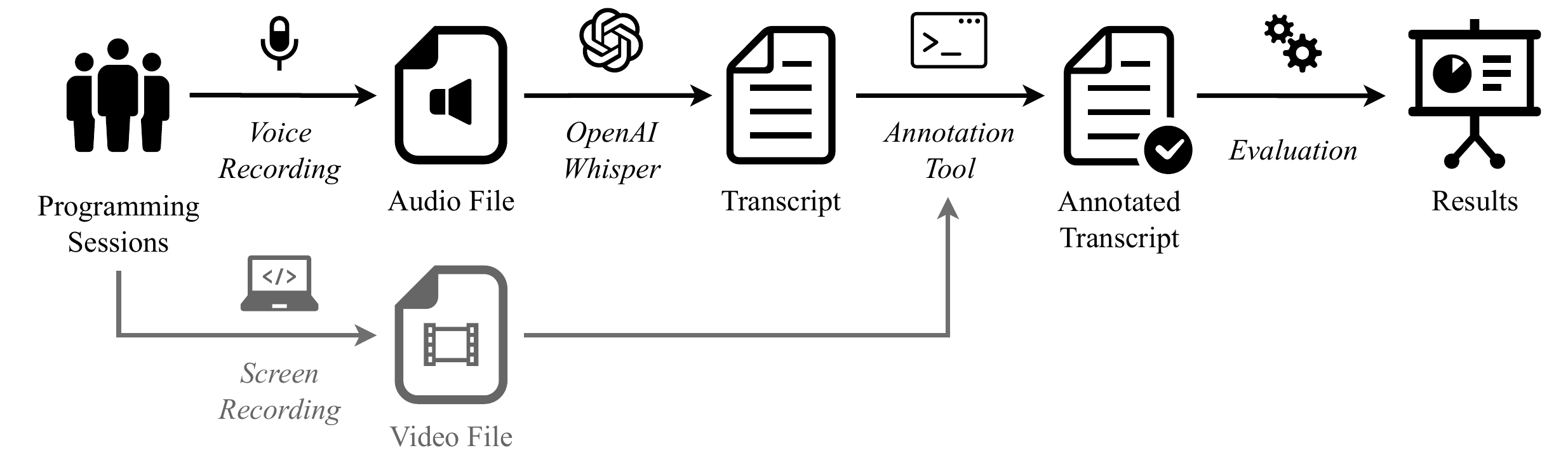}
    \caption{Overview of our data processing pipeline. While the annotated transcripts are primarily derived from voice recordings, screen recordings were consulted as needed to provide additional context.}
    \label{fig:eval:pipeline}
\end{figure*}

In our study, we examine differences in knowledge transfer between \a{humanhuman} and \a{humancopilot}.
We chose \a{copilot} because it is the most widely used coding assistant integrated into \acp{IDE}. 
In what follows, we outline our research goals, study setup, and data acquisition framework, along with details on data collection and processing procedures.

\subsection{Research Questions}

We first investigate the extent of conversations and interactions that facilitate \a{kt} in \a{humanhuman} and compare these results to the \a{humancopilot} setting to determine whether programmers using the AI assistant experience similar \a{kt} effects as in \a{humanhuman}.

\begin{rqquestion}{RQ\,1}
To what extend do the frequency, length, and depth of \a{kt} episodes differ between \a{humanhuman} and \a{humancopilot}?
\end{rqquestion}

We also examine how the two settings differ and align content-wise. We analyze topic type frequencies, identify the most discussed topic type, and assess whether this differs between settings.
The goal is to understand for which topic type \a{copilot} can transfer knowledge to developers and where it may fall short.

\begin{rqquestion}{RQ\,2}
How do the quality and diversity of \a{kt} episodes, including topic types and finish types, vary between \a{humanhuman} and \a{humancopilot}?
\end{rqquestion}

\subsection{Study Design}
We conduct a controlled experiment with two groups---a \a{humanhuman} group and a \a{humancopilot} group---and both groups complete the same programming task in \a{python}.
The \a{humanhuman} group consists of pairs working together without \a{copilot}, while the \a{humancopilot} group includes individuals using \a{copilot}.
Participants are randomly assigned to one of the two groups.

\suparagraph{Participants} We recruited students in the late stages of their bachelor's programs or the early stages of their master's programs.
All participants were either teaching assistants for foundational computer science courses or active members of the student council, ensuring a basic level of academic engagement and communication skills. 
In total, \numParticipantsTotal students took part in the study, resulting in \numSessionsPP \a{humanhuman} sessions and \numSessionsCopilot \a{humancopilot} sessions.

\suparagraph{Pre-test questionnaire} The study begins with a pre-test questionnaire, where
participants self-assess their general programming skills and \a{python} expertise along several dimensions using rating scales from 1 to 10, including a personal skill assessment relative to their peers~\cite{siegmund2014measuring}.
Participants then receive standardized verbal instructions covering project structure, time limit, and task location.
Participants in the \a{humanhuman} group are instructed to discuss with their partners, while those in the \a{humancopilot} group, having no human partner, are asked to think aloud, providing a spoken record of their thoughts for evaluation.

\suparagraph{Setting} Each participant works on a computer with an IDE set up, where the project is opened and necessary packages installed.
For the \a{humancopilot} group, an account is configured within the IDE to access \a{copilot}, and the plugin is activated. 
They are shown an introductory video on using \a{copilot}, while the 
\a{humanhuman} group receives an explanation of pair programming.
Both groups may use all resources, except AI assistants for the \a{humanhuman} group. All participants may ask questions before the session begins.

From this point, screen activity and conversations (or think-aloud processes) are recorded until submission or time expiration, resulting in audio and video files for evaluation.
Participants work on the task for 45 minutes, with no expectation to complete it.
This aspect is communicated to participants to avoid time pressure.

\suparagraph{Programming problem} 
Participants worked on a programming problem where certain features needed to be implemented within an existing codebase of approximately 400 lines including both \a{python} code and comments, distributed across 5 files. This setup ensures that participants have to develop an understanding of the broader context while working within a manageable scope. Specifically, participants received the source code for a password manager---a console application managing multiple user accounts, each with their own password entries.
After logging in, users can create, modify, view, and delete entries, which include a name, description, and password. 
The password manager stores data in a database, which is the focus of participants' implementation tasks, while the database schema and terminal interaction code are already provided. 

\suparagraph{Task} Participants' task is to complete functions marked with a \a{todoscode} comment, each with a provided signature and a doc string explaining the task.
These functions are confined to a single file and focus on database access (using SQLAlchemy\footnote{\url{https://www.sqlalchemy.org/}}) and building strings for console output. 
Two tasks involve string manipulation (e.g., building a formatted table of entries), while the remaining six focus on database interactions. 
Often, it is necessary to check conditions before modifying data (e.g., verifying if an object with a given name or id exists).

\subsection{Operationalization and Data Processing}

We evaluate knowledge transfer using our framework introduced in~\cref{sec:knowledgetransfer} to compare \a{humancopilot} and \a{humanhuman}.

To compare knowledge transfer, we analyze the distribution of quantities such as the number, depth, and length of episodes (see \cref{sec:knowledgetransfer}), which help us understand knowledge transfer dynamics.
Each new episode reflects a perceived knowledge gap, with knowledge transfer defined as efforts to close that gap by exchanging or building new knowledge.
To address RQ2, we examine the topic type (e.g., \textsc{Tool}, \textsc{Program}, \textsc{Bug}, \textsc{Code}, \ldots) and finish type (e.g., \textsc{Assimilation}, \textsc{Gave up}, \textsc{Trust}, \ldots) of these episodes. 

We collect data via voice and screen recordings during participants’ programming sessions. 
The data processing pipeline is shown in \cref{fig:eval:pipeline}.

\suparagraph{Transcription}
We transcribed the speech recordings using the automatic speech recognition tool \a{whisper}\footnote{\url{https://openai.com/index/whisper/}}.
\a{whisper} recognizes text in multiple languages, even within the same sentence.
This is useful since most participants spoke in German while using English for most technical terms.
We pre-processed the transcriptions to facilitate subsequent annotation by adding line breaks at sentence ends and removing filler words (e.g., \enquote*{okay}) that did not convey specific meaning.
It is important to note that during the annotation process, the original recordings remain essential, as vocal tone can convey emotions and irony more effectively than written text.

\suparagraph{Annotation} \label{annotation}
We developed a tool to annotate the transcript within the framework outlined in \cref{sec:knowledgetransfer}. 
This tool enables us to label each line of the transcript (one line equals one sentence) and to assign it to exactly one utterance.
We can also create episodes and assign utterances to them, editing episode attributes such as topic type (\textsc{Code}, \textsc{Bug}, ...) and finish type (\textsc{Trust}, \textsc{Lost sight}, ...).
In \a{pp} sessions, we recorded which programmer contributed to which lines, using  labels \enquote*{Speaker A} and \enquote*{Speaker B} for anonymity.
We annotated the transcripts from all \numSessionsTotal sessions, segmented them into utterances and episodes, and classified the utterances and episodes as described earlier.

\begin{figure*}[!h]
    \centering
    \begin{subfigure}{0.27\textwidth}
        \centering
        \includegraphics[scale=0.56]{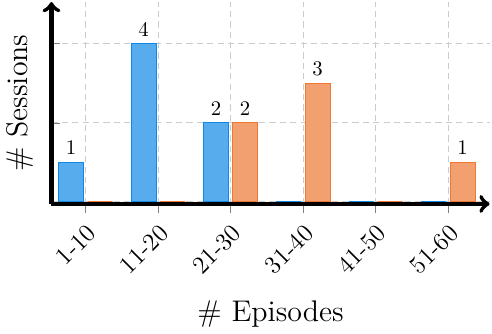}
    \end{subfigure}
    \hfill
    \begin{subfigure}{0.47\textwidth}
        \centering
        \includegraphics[scale=0.56]{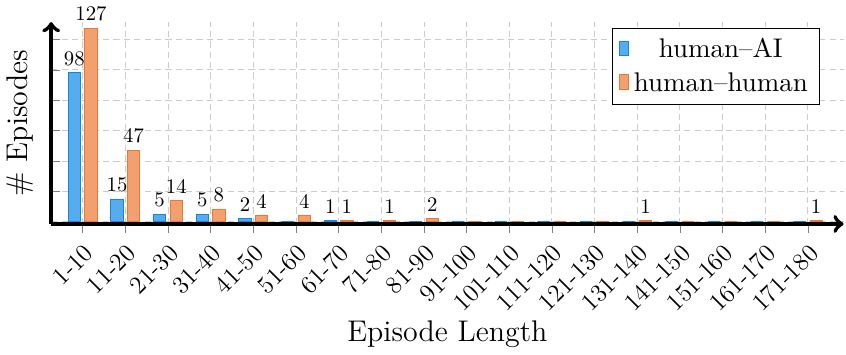}
    \end{subfigure}
    \hfill
    \begin{subfigure}{0.24\textwidth}
        \centering
        \includegraphics[scale=0.56]{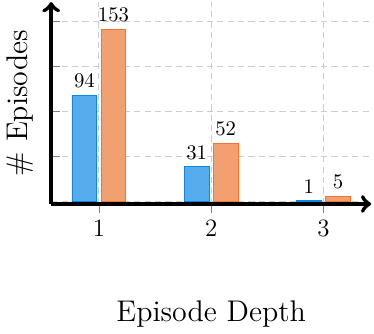}
    \end{subfigure}
    \caption{Comparison of episodes in terms of number (a), length (b), and depth (c).}
    \label{fig:eval:rqI}
\end{figure*}

\suparagraph{Segmentation}
We differentiate between monologues and dialogues, as \a{humanhuman} involves two people, while \a{humancopilot} features a single person articulating their thoughts as a result of the think-aloud process.
To segment the speech into utterances, we evaluate each spoken contribution determining whether it continues a previous utterance or begins a new one.
This decision is based on two factors: whether the contribution logically relates to the preceding one (i.e., addressing the same topic) and whether a noticeable pause or interruption occurs.
In dialogues, multiple consecutive utterances by one person can happen, but a new utterance begins when the other person speaks. 

\suparagraph{Classification} \label{classifciation}
For each utterance, we assess its contribution to \a{kt}.
In most cases, utterances related to \a{kt} contain an exchange of information between the two humans or between the human and \a{copilot}, as verified through the voice recording.
Additionally, we consider utterances that repeat information from external sources, such as documentation or Web-search results, as part of \a{kt} if it is apparent that it enhances a person's knowledge or understanding.
If the audio-stream alone lacks context, screen recordings were consulted.
From these utterances contributing to \a{kt}, we identify those indicating a need for knowledge or indicating a start of sharing knowledge, marking the start of an episode (\cref{episodedef}).

The episode continues until an utterance indicating one of the five discussed outcomes (e.g., \enquote{Yes, this makes sense,} classified as the finish type \textsc{Assimilation}).
While the audio transcript was sufficient in most cases, we referred to the original recordings when episode endings were unclear.
If an episode covers multiple topics but eventually returns to the original one, a new nested episode is recorded.
\suparagraph{Evaluation} For all statistical tests, we set the significance level to $\alpha = 0.05$. We analyze the number of episodes, initiated by a need for knowledge, as well as the length and depth of episodes. Additionally, we use annotations and classifications as described above to compare human-human and \a{humancopilot} settings. During classification, we also examine recurring session patterns, which are discussed further in \cref{sec:discussion}. In addition to what follows, we provide a comprehensive list of the statistical tests and their results in our replication package.

\begin{figure*}[!h]
    \centering
    \begin{subfigure}{0.36\textwidth}
        \centering
        \includegraphics[scale=0.6]{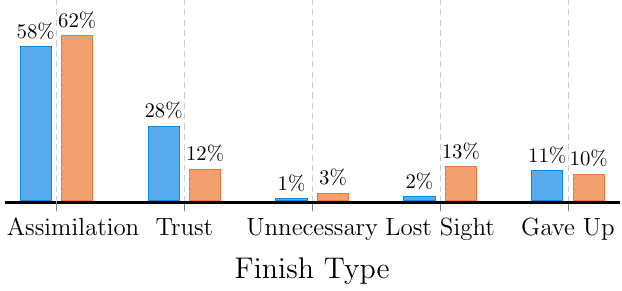}
    \end{subfigure}
    \hspace{-0.03\textwidth}
    \includegraphics[scale=0.6, trim=0 -3.5cm 0 0]{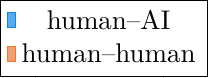}
    \hspace{-0.03\textwidth}
    \begin{subfigure}[t]{0.36\textwidth}
        \centering
        \includegraphics[scale=0.6]{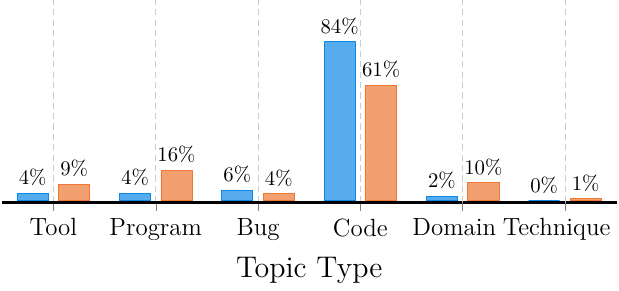}%
        
    \end{subfigure}
    \caption{Distribution of finish types (a) and topic types (b).}
    \label{fig:eval:rqII}
\end{figure*}

\section{Results}\label{sec:evaluation}
In this section, we evaluate and compare the knowledge transfer between traditional pair programming settings (\a{humanhuman}) and those involving a human paired with \a{copilot} (\a{humancopilot}) according to our two research questions.

\subsection{Demographics}
The following demographic characteristics provide essential context for our evaluation, serving as descriptive statistics for interpreting the study’s findings.
Most participants identified as male, with 84.2\,\% reporting as male and 15.8\,\% identifying as female. All participants were pursuing a computer science bachelor's or master’s degree.
With a mean value of \meanExpAbs out of 10, the self-indicated programming experience of our participants is rather high.
Furthermore, there is a positive Spearman's rank correlation between the self-indicated absolute programming experience and the self-indicated experience relative to the participants' peers, r\textsubscript{s}(\numParticipantsTotal)~=~\corrAbsRelExp. 
These experience measures are based on the work of \mbox{\citet{siegmund2014measuring}}.
A Mann-Whitney-U test did not yield any statistically significant differences for any of the five experience questions between the two groups.
However, the difference in experience with \a{copilot} stands out, with participants of the \a{humancopilot} group rating their experience with \a{copilot}, on average, at \meanExpCopilotCopilot (on a scale from 1 to 10), compared to an average rating of only \meanExpCopilotPP for the pair programming participants.
As a reminder, these differences were due to chance and all participants only answered the questions after being randomly assigned to a group.

\subsection{Characteristics of Knowledge Transfer Episodes}

We begin by examining the differences in frequency, length and depth of \a{kt} episodes guided by RQ1.

\suparagraph{Frequency of Knowledge Transfer}
\label{sec:evaluation:frequency}
As discussed in \cref{sec:methodology}, not every sentence in a conversation contributes to \a{kt}. Identifying significant differences in the number of episodes helps indicate how much and which parts of the conversation contribute to knowledge transfer.
To this end, we counted all episodes of both groups, regardless of their outcome or topic type.

In \a{humanhuman} sessions, there were in total \numEpTotalPP episodes, compared to \numEpTotalCopilot episodes in \a{humancopilot} sessions.
On average, pair programmers had \numEpMeanPP episodes per session and \a{copilot} users had \numEpMeanCopilot episodes per session. 
\cref{fig:eval:rqI}a contains the number of episodes per session across the two groups. 
The participants using \a{copilot} encounter \numEpFstQuCopilot to \numEpThdQuCopilot episodes, with a median of \numEpMedianCopilot.
In contrast, pair programmers ran into \numEpFstQuPP to \numEpThdQuPP episodes, with a median of \numEpMedianPP.
A Welch's t-test reveals that the mean difference between the number of episodes of the two groups is statistically significant, 
\welchwop{\wtNumEpT}{\wtNumEpDF}.

\suparagraph{Length and Depth of Episodes}
\label{sec:evaluation:length}
The lengths of episodes for both groups is shown in \cref{fig:eval:rqI}b.
On average, an episode of a \a{humanhuman} session spans \lenEpMeanPP utterances, compared to \lenEpMeanCopilot utterances in \a{humancopilot} episodes.
\changed{Notably, two extreme outliers in the \a{humanhuman} group could indicate that some episode endings may have been missed during annotation.
However, a manual inspection did not confirm this.
Instead, both outliers are top-level episodes that include smaller, nested episodes, thereby inflating their overall length.
This is to be expected, as our definition of episode length includes all utterances within an episode, even if they belong to nested episodes.
As a result, top-level episodes that encompass other episodes tend to be longer.}\@
The median is higher in the \a{humanhuman} group with \lenEpMedianPP utterances compared to \lenEpMedianCopilot utterances for the \a{humancopilot} group.
According to a Mann-Whitney U test, the distributions differ significantly. 
\cref{fig:eval:rqI}c shows the distribution of episodes depths. 
The highest occurring depth is 3 for both groups, and their distributions appear to be similar, with a Mann-Whitney U test revealing no significant difference between the two groups.
In both groups, most episodes are top-level episodes  (i.e., a depth of 1).
Thus, they were caused by a need for knowledge that did not result from another foregoing episode.
Only one quarter of all episodes are stacked on top of a different one, and a negligible small share of $1$--$2\,\%$ of episodes is stacked on two episodes.

\subsection{Characterizing Knowledge Transfer Dynamics}
We continue with RQ2, that is, the question of differences in topic type and outcome of knowledge transfer episodes.

\suparagraph{Content of Episodes}
The distribution of episode topic types is shown in \cref{fig:eval:rqII}b.
The \yaxis shows the percentage of episodes with the bars placed at the respective topic type on the \xaxis.
As the number of episodes is different for the two groups, we chose relative numbers for the plot.
A $\rchi^2$-test reveals a significant difference between the two distributions. 

Across both groups, the topic type \textsc{Code} occurs most frequently.
However, it occurs significantly more in the \a{humancopilot} group, accounting for $84\,\%$ of all episodes, with other topic types only occurring in at most $6\,\%$ of all episodes.
In both groups, there was little discussion about abstract programming paradigms (with $0\,\%$ and $1\,\%$ of episodes having topic type \textsc{Technique}).

\suparagraph{Outcomes of Episodes}
We analyze the outcomes of knowledge transfer episodes with regard to finish types to assess the quality of \a{kt}. This is the most crucial aspect because knowledge transfer is only useful to the customer if they can process and understand the information conveyed to them.

\begin{figure}[H]
    \centering    
    \includegraphics[width=0.75\linewidth]{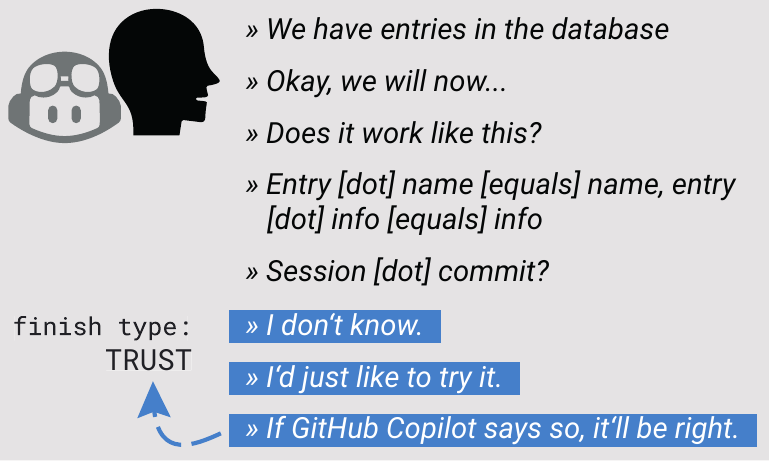}
    \caption{Episode with finish type \textsc{Trust}. Guillemets mark single utterances. Bracketed text added for context.}
    \label{fig:eval:epepisodeepisodetrust}
\end{figure}

\Cref{fig:eval:rqII}a shows how finish types are distributed per group.
On the \yaxis, we see the percentage of episodes with the respective finish type. 
According to a $\rchi^2$-test, the distributions differ significantly.
Both groups tend to finish most episodes successfully, with a
12\% difference between the two settings in successful \a{kt} episodes(finish type:  \textsc{Assimilation} and \textsc{Trust}).
More interesting is the discrepancy in the category \textsc{Trust}, with \a{copilot} users achieving this finish type more than twice as often than users in  \a{humanhuman}.
Upon closer examination of episodes finish type \textsc{Trust}, we observe that in many \a{copilot} sessions, programmers tend to accept the assistant’s suggestions with minimal scrutiny, relying on the assumption that the code will perform as intended. 
\Cref{fig:eval:epepisodeepisodetrust} shows an example of this happening.
This issue does not come up in \a{pp} sessions so much, indicating that pair programmers rather pursue problems until the need for knowledge is satisfied.

\begin{figure}[H]
    \centering    
    \includegraphics[width=0.75\linewidth]{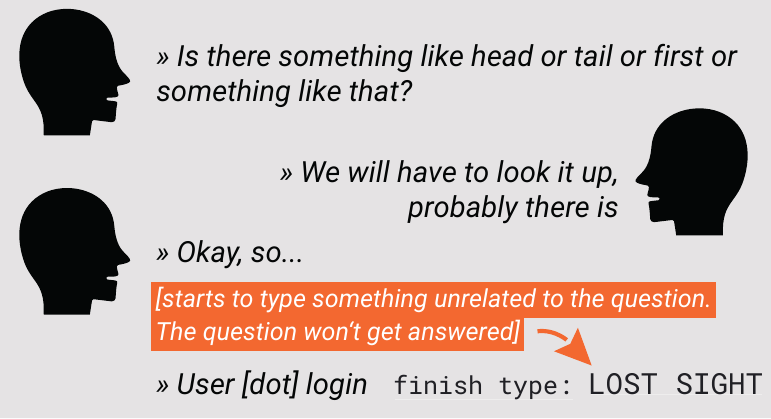}
    \caption{Episode with finish type \textsc{Lost sight}. Guillemets mark single utterances. Bracketed text added for context.}
    \label{fig:eval:epepisodeepisodelostsight}
\end{figure}

Another category with a huge discrepancy is \textsc{Lost sight}. 
\Cref{fig:eval:rqII}a shows that this almost does not occur in the \a{humancopilot} group, but one eighth of episodes of the \a{humanhuman} group are interrupted by an unrelated issue (see~\cref{fig:eval:epepisodeepisodelostsight} for an example).

By performing $\rchi^2$-tests to compare the occurrence frequencies of individual finish types, we found significant differences between the two groups for the \textsc{Trust} and \textsc{Lost sight} finish types, while we could not find significant differences for the other three finish types.


\section{Discussion }\label{sec:discussion}

Even the most experienced programmer will reach a point in their day-to-day programming where some additional knowledge would be beneficial. In our case, the programming scenario involved a fairly self-contained programming task, during which we already observed plenty of knowledge transfer. How this knowledge transfer differs between traditional human pair programming and the AI-assisted approach was the core of our investigation that we will discuss in the following.

\subsection{Findings}

Regarding RQ1, our initial objective was to determine whether knowledge transfer occurs within \a{humancopilot} and how the frequency, length, and depth of knowledge transfer episodes in \a{humancopilot} compare to \a{humanhuman}. 

We followed the definition by \mbox{\citet{zieris2014knowledge}} and defined knowledge transfer as \textit{any attempt} to close a perceived knowledge gap through the exchange or creation of new knowledge. 
These attempts to close knowledge gaps indeed occurred in both scenarios.
The average number of episodes per session differs significantly between the groups, which may indicate either a greater need for knowledge exchange or simply a lower threshold for initiating these interactions in \a{humanhuman}. This suggests that individuals may be more eager to engage with each other.

In a manual investigation of the transcripts, we found many relatively short question-and-answer episodes in \a{humanhuman} sessions that do not emerge in \a{humancopilot} sessions, supporting our assumption. 
These question-and-answer episodes usually follow a pattern in which the customer casually poses a question that the partner can answer quickly. 
\changed{
This does not occur when using the \a{copilot} assistant as, at the time the study was conducted, the only way to interact with the assistant was by waiting for a suggestion. Although it was technically possible to prompt the assistant through comments, this method was unintuitive, making it unlikely that participants would pose minor questions.}

\begin{rqquestion}{Finding\,1}
\A{kt} occurs in both human-human and \a{humancopilot} settings, revealing knowledge gaps and efforts to bridge them.
However, in \a{humanhuman} settings, individuals tend to engage more actively with each other, often involving relatively short question-and-answer episodes.
\end{rqquestion}

We examined the length of episodes by counting the number of utterances. The median episode length is significantly higher in the \a{humanhuman} group compared to the \a{humancopilot} group. 
This difference might be expected, as a dialogue between two human partners naturally involves more exchanges than a monologue where one person thinks aloud. However, one might also assume that discussions in a pair programming setting would lead to faster conclusions--after all, that is one of the reasons for pair programming--but our study did not observe this effect. Without the proper context or goal in mind, it is difficult to conclude which method is superior. Shorter discussions might work well when quick decision-making is needed, while longer ones allow for thorough analysis and idea development.

Episodes with higher depths occur when the participant in the customer role requires knowledge on a topic distinct from the current episode's topic. In both groups, most episodes remain top-level, with a depth of 1. Deeper and longer episodes may suggest that certain explanations are unclear, as additional context often becomes necessary to clarify related topics. Interestingly, there is no significant difference in the distribution of episode depths.

\begin{rqquestion}{Finding\,2}
While \a{humanhuman} interactions tend to produce longer episodes, the depth of episodes remains similar across both groups. This suggests that the need for additional contextual explanations is comparable.
\end{rqquestion}

Addressing RQ2, we analyzed how the quality and diversity of knowledge transfer episodes in pair programming with \a{copilot} compare to \a{humanhuman} by examining the frequency of different topic types and finish types in both groups.

For both settings, most episodes were assigned the \textsc{Code} topic type. During manual investigation, we found that this often included episodes in the \a{humancopilot} setting where a programmer creates code with the help of the assistant and verifies (verbally) that the code actually fulfills its purpose. 
This form of \a{kt} was named a \enquote{co-production} by \mbox{\citet{zieris2014knowledge}}.
Originally, for \a{humanhuman}, this defines episodes where the programmers work together on building new knowledge (e.g., discovering a bug and its cause) \cite{zieris2014knowledge}.
This again shows that in such scenarios, similar to the \a{humanhuman} setting, the programmer actively builds new knowledge by trying to understand the code presented by \a{copilot}. 

When comparing the topic types of knowledge transfer, we observe that there is a broader distribution in the \a{humanhuman} in which other topic types besides \textsc{Code} occur more often than in the settings where \a{copilot} is involved. We hypothesize that this is due to episodes with topic types such as \textsc{Program} or \textsc{Domain} being more likely to start with casual conversation than a concrete question, making them more likely to occur with a human partner. \changed{Additionally, the suggestions made by \a{copilot} are typically closely tied to the project's source code, making the emergence of unrelated topics unlikely.}{}

Regarding less prevalent topic types, there are only few episodes on the topic type \textsc{Program} (i.e., thus, regarding syntax or semantic of the programming language).
This can be explained by the high \a{python} experience that participants had, on average.
Also, participants rarely discussed programming concepts (topic type \textsc{Technique}). 
They might not have needed to address them because the design of the project as well as the conditions under which they were supposed to work were all given.
The relative rarity of the \textsc{Bug} topic type is likely because, when errors occur, discussions tend to focus less on their cause and more on finding a solution. 
In some cases, this also led to the code in question being deleted and rewritten, which meant that the reference to the error was usually removed.

\begin{rqquestion}{Finding\,3}
In general, \a{humancopilot} sessions focus more on the \textsc{Code} topic type.
Conversely, knowledge transfer in \a{humanhuman} involves a broader range of topic types, beyond mere code-related discussions. 
\end{rqquestion}

Regarding the initiation of episodes, Zieris and Prechelt~\cite{zieris2020qualitative} differentiate between pull mode, where one person actively seeks information from their partner to close a knowledge gap, and push mode, where one person shares information based on the perception that their partner might need it.
We cannot directly infer push mode episodes coming from \a{copilot}, since all episodes in the \a{humancopilot} group are purely comprised of utterances by a human participant. However, we noticed some interesting patterns, which show that \a{copilot} can sometimes convey knowledge subtly, even without an explicit prompt \changed{or preceding comment}.
Most participants in the \a{humancopilot} group did not have any experience with the database API \textsc{SQLAlchemy}, which was used in the project.
Hence, most forgot to add a commit statement after doing database transactions in the code. 
In many cases, \a{copilot} suggested that adding such commit statement is crucial for the program to function correctly. 
Upon reading the suggestion in one function, participants realized that they were to add this statement in every other function that modifies the database as well.
This elegantly shows that \a{kt} does not exclusively happen \enquote{on request} but can be rather subtle and unsolicited. Programmers may learn something not only from another human but also via an AI coding assistant being proactively pushing suggestions.

\begin{rqquestion}{Finding\,4}
\a{copilot} can also convey critical information subtly.
However, we suspect the transfer of subtle, tacit knowledge to occur less frequently in real-world settings where \a{copilot} lacks access to company-specific information and domain expertise.
\end{rqquestion}

In \a{humancopilot}, we observe a significantly higher share of successful knowledge transfer episodes (finish types \textsc{Assimilation} and \textsc{Trust}) than in \a{humanhuman} (see Figure \ref{fig:eval:rqII}). 
A key reason for this finding appears to be the significantly higher probability of distraction in the human--human setting (finish type \textsc{Lost sight}) compared to \a{humancopilot}.

In addition to the low likelihood of distraction, we find that knowledge transfer with the finish type \textsc{Trust} occurs much more frequently in \a{humancopilot}. Programmers tend to accept the assistant’s suggestions with minimal scrutiny, relying on the assumption that the code will perform as intended.

\begin{rqquestion}{Finding\,5}
Programmers tend to accept the assistant's suggestions with little critical review, often trusting that the code will work as expected---a tendency seen far more frequently than in the human--human setting. Conversely, conversations initiated with \a{copilot} are less likely to be aborted due to programmers getting distracted.
\end{rqquestion}

\subsection{Threats to Validity}
\textbf{Internal Validity.} 
Think-aloud protocols may not capture all participant thoughts, as verbalization can be incomplete or cognitively demanding, potentially reducing performance or slowing task completion \cite{charters2003think}.
As a result, transcripts---especially in the \a{humancopilot} group---may miss knowledge transfer episodes, suggesting our findings represent a lower bound. On the other hand, thinking aloud may also encourage structured thinking and improve task focus.

Additionally, data was annotated by a single person, introducing the risk of individual bias in the classification of knowledge transfer episodes. To mitigate this, we automated large parts of the processing and manually cross-checked a random sample of annotations with four additional reviewers.

\textbf{External Validity.} 
Participants had solid programming experience and reflected a typical gender distribution of computer science students. To isolate the effect of \a{humancopilot} on knowledge transfer, we controlled for factors like domain knowledge, peer relationships, and organizational culture. While this strengthens internal validity~\cite{siegmund2015views}, it may limit ecological validity, as collaborative behavior can depend on personal traits and real-world dynamics. Thus, generalization to more diverse populations should be explored.

The observed tendency to trust \a{copilot} may result from situational factors like time constraints, lack of consequences for errors, or the perceived effort of manual verification.

We used Python for the programming task, which may influence \a{copilot}’s performance~\cite{nguyen2022empirical}. However, we have no indication that our core findings on knowledge transfer would not generalize to other commonly used languages.

\textbf{Construct Validity.} Episode length was operationalized by utterance count, not time, due to missing timestamps. 
This approach does not reflect processing time, but emphasizes the amount of content exchanged, making it more robust against variation in speaking pace.
Apart from that, we drew on existing frameworks to examine the relevant constructs.

\section{Conclusion}\label{sec:conclusion}

The promise of AI coding assistants evolving into true \emph{AI pair programmers} has been a compelling narrative.
Our findings show that, while they support knowledge transfer and offer proactive guidance, AI assistants 
do not yet replicate the diversity of human collaboration.

In our study with \numParticipantsTotal participants, knowledge transfer occurred both in \a{humanhuman} and with \a{copilot}. Notably, \a{copilot} can subtly remind developers of important details, such as committing database changes, that might otherwise be overlooked. This unintentional form of knowledge transfer was previously assumed to be limited to human collaboration.
\A{humanhuman} enables spontaneous interactions but also increases the risk of distraction. In contrast, \a{kt} with \a{copilot} is less likely to be aborted, yet suggestions are often accepted with less scrutiny. This highlights the need for further research---technical and social---to understand trusting behavior and to develop mechanisms that encourage critical evaluation. 

Our results suggest that combining AI assistants with traditional pair programming can balance efficiency with richer knowledge exchange.
For instance, senior developers can offer contextual insights beyond AI capabilities, but may benefit from \a{copilot}'s support with repetitive tasks.

\section*{Data Availability}\label{sec:data}

For transparency and reproducibility, we share all study artifacts online\footnote{\url{https://anonymous.4open.science/r/knowledgeTransferCopilot-BF28/README.md/}}.

\def\bibfont{\footnotesize}
\bibliographystyle{IEEEtranN}
\bibliography{bib}
\end{document}